\documentclass{aa}  
\usepackage{graphicx}
\usepackage{txfonts}
\usepackage{lineno}

\newcommand{\eboss}{E-BOSS catalogue}  
\newcommand{\fermi}{\emph{Fermi}}
\newcommand{\lat}{\emph{Fermi}-LAT}
\newcommand{\eflux}{$\epsilon_\gamma F(\epsilon_\gamma ) $}
\newcommand{\zetaoph}{$\zeta$~Ophiuchi}
\newcommand{\bd}{BD$+43\degr3654$}

\begin{document} 

   \title{Systematic search for high-energy gamma-ray emission from bow shocks of runaway stars}
	\titlerunning{Search for high-energy stellar bow shock emission }
\author{
A.~Schulz$^{1}$ \and 
M.~Ackermann$^{1}$ \and 
R.~Buehler$^{1}$ \and 
M.~Mayer$^{1}$ \and 
S.~Klepser$^{1}$
}

\institute{
\inst{1}~Deutsches Elektronen Synchrotron DESY, D-15738 Zeuthen, Germany\\ 
\email{anneli.schulz@desy.de} \\
}
   \date{Received ?; accepted ?}

  \abstract
   {It has been suggested that the bow shocks of runaway stars are sources of high-energy gamma rays (E\,>\,100\,MeV). Theoretical models predicting high-energy gamma-ray emission from these sources were followed by the first detection of non-thermal radio emission from the bow shock of \bd and non-thermal X-ray emission from the bow shock of AE Aurigae.}
   {We perform the first systematic search for MeV and GeV emission from 27 bow shocks of runaway stars using data collected by the Large Area Telescope (LAT) onboard the \textit{Fermi Gamma-ray Space Telescope (Fermi)}.}
   {We analysed 57 months of \lat~data at the positions of 27 bow shocks of runaway stars extracted from the Extensive stellar BOw Shock Survey catalogue (E-BOSS). A likelihood analysis was performed to search for gamma-ray emission that is not compatible with diffuse background or emission from neighbouring sources and that could be associated with the bow shocks.}
   {None of the bow shock candidates is detected significantly in the \lat~energy range. We therefore present upper limits on the high-energy emission in the energy range from 100\,MeV to 300\,GeV for 27 bow shocks of runaway stars in four energy bands. For the three cases where models of the high-energy emission are published we compare our upper limits to the modelled spectra. Our limits exclude the model predictions for \zetaoph~by a factor $\approx 5$.  }
   {}

   \keywords{ Radiation mechanisms: non-thermal -- Gamma rays: ISM; stars -- Stars: early-type -- Stars individual: \zetaoph   }
	\maketitle

\section{Introduction}
Runaway stars with strong winds can produce bow shocks in the surrounding interstellar medium (ISM) when moving supersonically with respect to the material. The runaway stars sweep up the ISM in the direction of motion, and arc-shaped features develop ahead of the stars. Thermal emission from many bow shocks of runaway stars has been detected at infrared wavelengths. The mid- to far-infrared radiation originates in the dust that is swept up by the supersonic movement of the stars through the ISM and heated by the stellar radiation, as well as by the radiation from the shocked gas. Stellar bow shocks were discovered by \cite{vanBuren88} using data from the \textit{Infrared Astronomical Satellite} (IRAS), and the first survey was performed by \cite{vanBuren95}. 
   
The most recent survey of bow shocks of runaway stars is the Extensive stellar BOw Shock Survey catalogue \citep[ ][E-BOSS]{eboss}. It summarizes a systematic search for bow shocks around runaway OB stars in the newest infrared data releases, mainly using data from the Midcourse Space eXperiment (MSX) and the Wide-field Infrared Survey Explorer (WISE). Their search around 283 early-type stars, selected to be closer than 3\,kpc, results in a sample of 28 bow shock candidates. Bow shocks can thus be detected around roughly 10\% of the runaway OB stars. \cite{eboss} do not find any correlation between the detection of a bow shock and either stellar mass, age or position.

Introducing a non-thermal emission model, \cite{bd_benaglia} suggest that bow shocks are emitters of high-energy gamma rays (HE, E\,>\,100\,MeV). Moreover, it was shown that the emission could be detectable by current gamma-ray experiments \citep{Valle_model_zeta}.

Bow shocks can accelerate particles up to relativistic energies via Fermi shock acceleration.

\cite{bd_benaglia} have detected non-thermal radio emission from the bow shock of \bd, which is produced by accelerated electrons that emit synchrotron radiation. The same electrons upscatter photons from the stellar and dust photon fields via the inverse Compton process, which leads to high-energy gamma-ray emission. The search for an X-ray counterpart of the bow shock of \bd~performed by \cite{terada_bd} using a 99\,ks exposure with \textit{Suzaku} only resulted in upper limits. They measure an enhanced X-ray count rate in the bow shock region, but taking the systematic errors on the non X-ray and cosmic-ray background into account, they argue that the X-ray count rate in the bow shock region is compatible with that of the background region.

The first detection of non-thermal X-ray emission from a bow shock produced by a runaway star was recently claimed by \cite{ae_aurigae_LopezSantiago} for AE Aurigae (HIP 24575). Although the XMM-\textit{Newton} data do not allow distinguishing between a very hot thermal and a non-thermal origin, the latter seems more likely for two reasons: there is no counterpart in the infrared and optical wavelengths for the putative thermal source, which could be a foreground or background stellar object, and the temperature would have to be extremely high. The good spatial correlation of the X-ray and IR emission from the bow shock also promotes the bow shock hypothesis.

The first potential detection of a bow shock from a runaway star in high-energy gamma rays was published by \cite{Valle_pulsar}. The bow shock of HD 195592, listed in the \eboss~as HIP 101186, is spatially coincident with the \fermi~source 2FGL~J2030.7+4417 \citep{2fgl}. Under some energetic assumptions \cite{Valle_pulsar} conclude that 2FGL~J2030.7+4417 might be associated with the bow shock from HIP 101186. However, this \fermi~source has been identified as a gamma-ray pulsar by \cite{Pletsch}. In the Second \fermi~Large Area Telescope Catalog of Gamma-ray Pulsars \citep{2013ApJS..208...17A}, it is listed among the pulsars with no significant off-peak emission. The absence of off-pulse emission is a clear indicator that the observed LAT photons predominantly originate in the pulsar. A possible gamma-ray signal from the bow shock of HIP 101186 is below the current sensitivity threshold of the LAT. Gamma rays from the bow shock might be detected with deeper LAT observations or by observations with future instruments like the Cherenkov Telescope Array that feature better angular resolution than the LAT. 

In this paper we describe the \lat~observation and data analysis of 27 bow shock candidates in Section\,2. The results of the analysis of 57 months of \lat~data are presented in Section\,3, which also includes a comparison of the calculated \lat~upper limits with published model predictions for the three cases where these are available. We conclude with some implications of the non-detections and a short look at which instruments might be able to detect or further constrain the high-energy emission from the bow shocks of runaway stars.   

\section{\lat~observation and data analysis}
We used the 28 bow shock candidates listed in the \eboss~as the basis for our search, with the exception of HIP 101186. Since emission from the bow shock cannot be separated reliably from the emission of the bright gamma-ray pulsar \citep[reported in][]{Pletsch}, it is omitted in this study. 

The \fermi~Large Area Telescope (LAT) is a pair conversion telescope onboard the \fermi~\textit{Gamma-ray Space Telescope}. It was launched in 2008 and surveys the gamma-ray sky in the energy range from 20$\,$MeV to over $300\,$GeV. Details of the instrument are described in \cite{performance2009}, while the on-orbit performance of the telescope is described in \cite{performance_2012}. The point-spread function (68$\%$ containment angle, PSF) of the \lat~decreases strongly with increasing energy: from 6\degr~at 100$\,$MeV to 0.25\degr~at 10$\,$GeV.   

We analysed the \lat~data from the beginning of scientific operations on 2008 August 4 to 2013 May 2 and searched for emission from the 27 bow shock candidates, selected as explained above. We select events classified as photons in the P7SOURCE event classification \citep{performance_2012} in a region of interest (ROI) of 15\degr~radius around the potential source with energies from 100\,MeV to 300\,GeV. To reduce the contamination by gamma rays produced by the interactions of cosmic rays with the upper atmosphere, we excluded photons arriving from angles larger than 95\degr~with respect to the zenith. In addition, time intervals in which the \lat~rocking angle exceeds 52\degr~were excluded in the analysis. 
We used the \fermi~Science Tools (v9r29p0)\footnote{\url{http://fermi.gsfc.nasa.gov/ssc/data/analysis/software/}} to analyse the photon data binned in energy and arrival direction. The spectra of the potential sources are calculated with a binned likelihood fit in 30 energy bins over a $20\degr \times 20\degr$~region on a grid of 0.1\degr, which is centred on the position of the runaway star. We used the P7SOURCE\_V6 Instrument Response Functions and corresponding models for the Galactic and isotropic diffuse emissions (\textit{gal\_2yearp7v6\_v0.fit} and \textit{iso\_p7v6source.txt}\footnote{\url{http://fermi.gsfc.nasa.gov/ssc/data/access/lat/BackgroundModels.html}}).  

\begin{figure}
 \resizebox{\hsize}{!}{\includegraphics{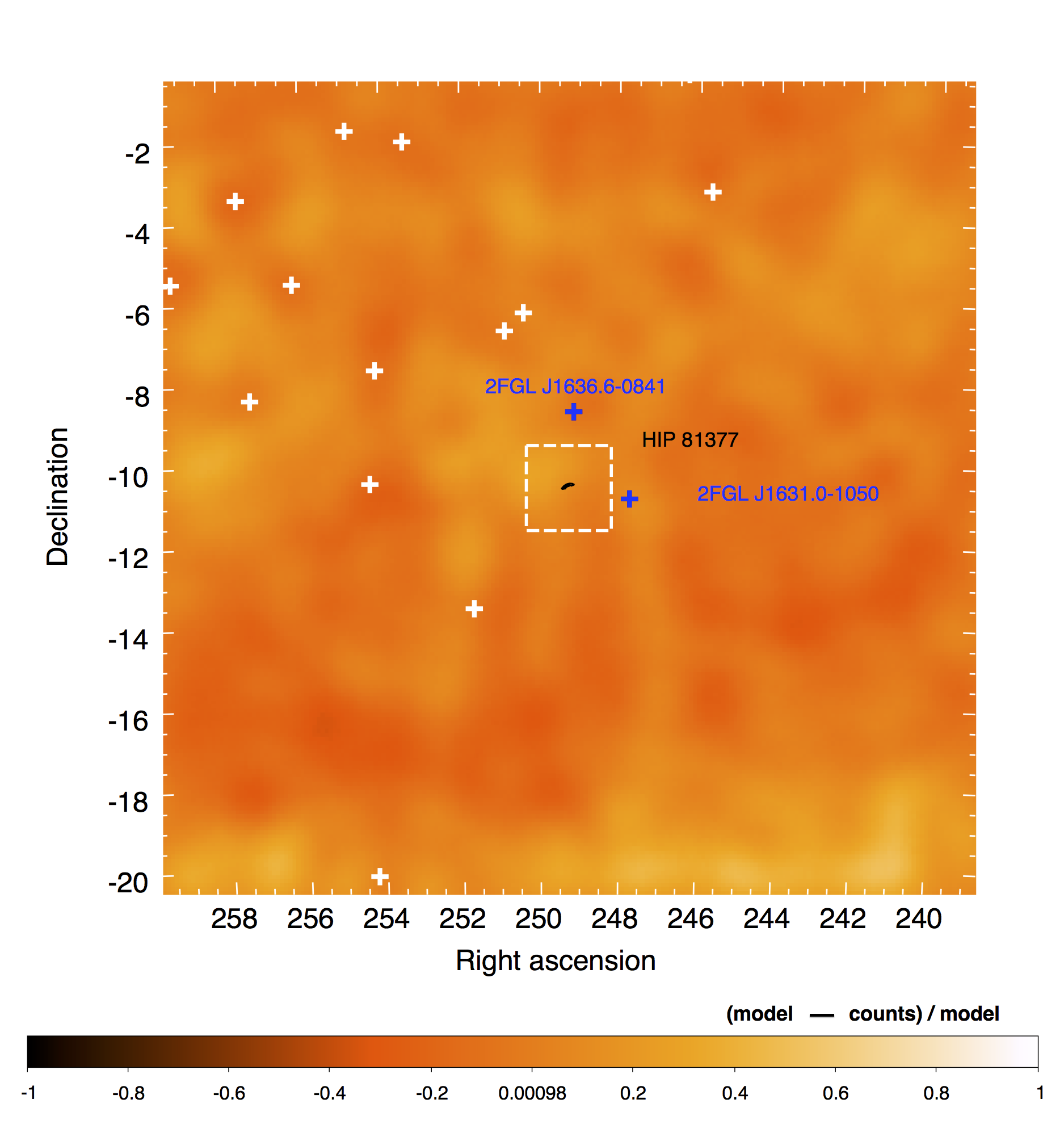}}
  \caption{Residual count map  in the energy range from 100\,MeV to 300\,GeV on a 20\degr~square around HIP 81377 with a bin size of 0.1\degr. The map has been smoothed with a Gaussian kernel of 1\degr. Blue crosses denote sources with free spectral parameters in the fit. White crosses denote positions of 2FGL sources with spectral parameters fixed to the catalogue values in the fit. The WISE contours (22 micron) are shown in black. The white dashed box depicts the size of the TS map, shown in Fig.\,\ref{Fig:tsmap_HIP81377}.}
  \label{Fig:resmap_HIP81377}
\end{figure}

In addition to the diffuse sources, the likelihood fit needs models of any gamma-ray point sources in or near the ROI. The first step to create the input model is to take all sources listed in the second \lat~catalogue \citep[][2FGL]{2fgl} into account that are closer than 17\degr~to the potential source. We hold the flux and spectral index fixed for sources more than 3\degr~away from the bow shock at the values obtained from the 2FGL. The spectrum parameters of closer sources were redetermined in our likelihood fit. We used the same spectrum parametrization as in the catalogue.  
Since our data set is much larger than the one used for the 2FGL, we calculated residual count maps of the analysed regions by subtracting the observed counts from the expected counts (with respect to our fitted model) and dividing by the model map. If bright residuals appear at the position of a 2FGL source that was held fixed in the first fit, we additionally released the spectral parameters of this source during a second iteration of the likelihood fit, independent of the distance to the potential source. An example of such a residual count map including events from 100\,MeV to 300\,GeV is shown in Fig.\,\ref{Fig:resmap_HIP81377}.

For one ROI the residual map shows bright emission at a place with no associated 2FGL source: this case is HIP 78401, where a bright residual is at the position of a recently published source \citep{fava}, namely Fermi J1532--1319. Consequently, an additional source with a power-law spectral shape is included in the model. Its spectral parameters are free to vary during the likelihood fit. 
To assess the significance of the gamma-ray signal of a potential source, we calculated the maximum-likelihood test statistic (TS) following \cite{1996ApJ...461..396M}: $\mathrm{TS} = -2\cdot \log ( \mathcal{L}_{\mathrm{max},0} / \mathcal{L}_{\mathrm{max},1}) $, where $\mathcal{L}_{\mathrm{max},0}$ and $\mathcal{L}_{\mathrm{max},1}$ are the maximum likelihood for a model without the source (null hypothesis) and with the additional source, respectively.
Applying this relation, we calculated TS maps on a 0.1\degr~grid in a square with 2\degr~edge length centred on the bow shock (depicted in Fig.\,\ref{Fig:resmap_HIP81377})  and search for localized excesses in the TS map. We add a source to the model if the significant emission does not coincide with the bow shock position. 

\begin{figure*}
  \includegraphics{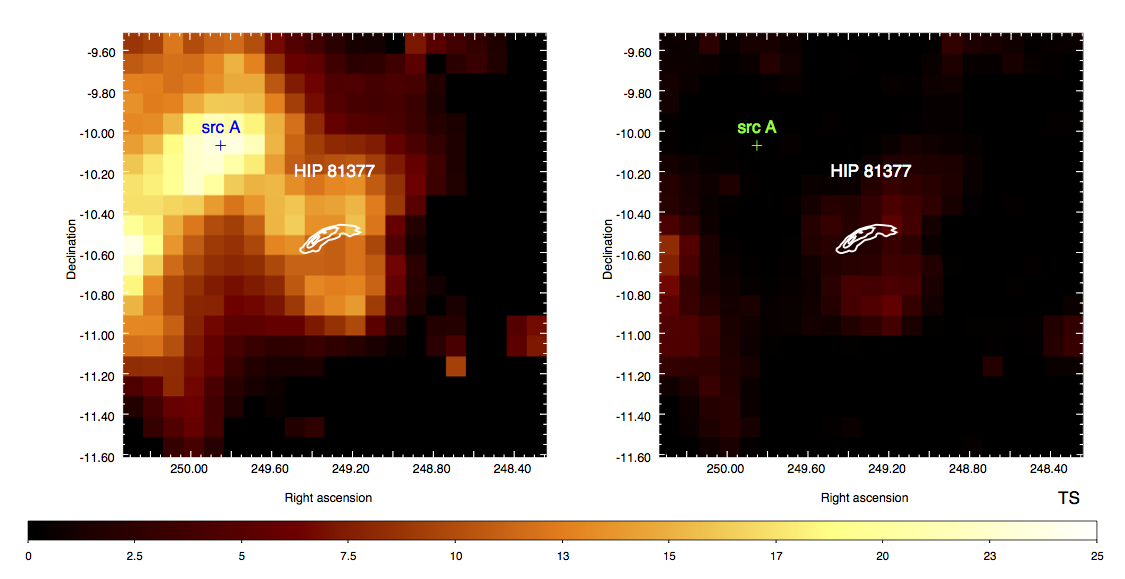}
  \caption{TS map centred on the position of \zetaoph~(HIP 81377), on a 2\degr~square with a bin size of 0.1\degr. The blue (green) cross indicates the position of the additional source A on the left (right) map. White contours show the bow shock of \zetaoph~observed in infrared at $22\,\mu$m by WISE. On the left side, only 2FGL sources are included in the fitted model, in the right map the additional source A is also included.}
  \label{Fig:tsmap_HIP81377}
\end{figure*}

This procedure is similar to the one pursued in the 2FGL catalogue, and it results in two additional sources in the field of view of HIP 32067 and one additional source for HIP 38430 and HIP 81377. Power laws are used to model the spectra of the additional sources at the positions RA(J2000) = 102.01\degr, Dec(J2000) = 6.83\degr and RA(J2000) =  100.74\degr, Dec(J2000) = 5.35\degr~for HIP 32067. For HIP 38430 the additional source is at the position RA(J2000) =  118.52\degr, Dec(J2000) =  $-$26.83\degr~and for HIP 81377 at RA(J2000) = 249.85\degr, Dec(J2000) = $-$10.072\degr. The TS maps for the latter source are shown in Fig.\,\ref{Fig:tsmap_HIP81377}, where the left-hand map shows the analysis with only the 2FGL sources included in the model. This map leads to the addition of "source A" in the model for the subsequent fit. The TS map with source A included in the model is shown on the right-hand side of Fig.\,\ref{Fig:tsmap_HIP81377}.  

The morphology of the potential gamma-ray emission from the bow shock of a runaway star is expected to be similar to the infrared emission. Given the above-mentioned characteristics of the \lat~PSF, we decided to use a template if either the size of the bow shock is larger than 0.3\degr~or the distance from the bow shock to the star is larger than 0.1\degr. This leads to four bow shock candidates, which we consider to be potentially extended for \lat: HIP 22783, HIP 78401, HIP 81377, and HIP 97796. In all other cases we search for gamma-ray emission from a point-like source at the location of the runaway star. The WISE templates\footnote{obtained via \url{http://skyview.gsfc.nasa.gov}} ($22\,\mu$m) are processed to be used with the \fermi~Science Tools: after masking regions immediately outside the bow shock, each template is normalized to 1.

\section{Results and discussion}
We calculated TS values for all 27 bow shock candidates assuming a power-law spectral model. All potential sources have TS values below 10; i.e., no significant detection of gamma-ray emission from a bow shock has been made (see Table\,\ref{table:UpperLimits}). We used a Bayesian approach \citep[following][]{Helene1983319} to calculate upper limits on the flux in the energy range from 100\,MeV to 300\,GeV in four energy bins, assuming a power-law spectrum of gamma-ray emission ($dN/dE = N_0 {(E/E_0) }^{-\alpha}$) with photon index $\alpha$=2 in each band. The $95\%$ confidence-level, gamma-ray flux upper limits in four energy bins (equally spaced in logarithmic energy) are presented in Table\,\ref{table:UpperLimits}.

\begin{table*}
\caption{$95\%$ confidence-level, gamma-ray flux upper limits for bow shocks of runaway stars. }              
\label{table:UpperLimits}      
\centering                                     
\begin{tabular}{l |c |c | c |c | c| c |c}          
\hline\hline         
Star & $l$ & $b$&  TS & \multicolumn{4}{c}{\eflux [$10^{-6}\,$MeV\,cm$^{-2}$\,s$^{-1}$]} \\
 & [\degr~] &[\degr~] & &$0.1 - 0.74$ & $0.74 - 5.5$ & $5.5 - 41 $ & $41 - 300$ [GeV] \\
\hline 
HIP 2036        & 120.9137 & +09.0357 &  0 & 0.61 & 0.13 & 0.30 & 1.44 \\
HIP 2599        & 120.8361 & +00.1351 & 1& 0.99 & 0.55 & 0.39 & 0.99 \\
HIP 11891       & 134.7692 & +01.0144 & 7 &1.20 & 0.60 & 0.79 & 1.61 \\
HIP 16518       & 156.3159 & $-$16.7535 & 0 & 0.93 & 0.22 & 0.21 & 1.29 \\
HIP 17358       & 150.2834 & $-$05.7684 & 0 & 0.56 & 0.24 & 0.24 & 1.26 \\
HIP 22783$^*$       & 144.0656 & +14.0424 & 0 & 0.82 & 0.15 & 0.19 & 0.72 \\
HIP 24575       & 172.0813 & $-$02.2592 & 0 &0.21 & 0.19 & 0.28 & 0.96 \\
HIP 25923       & 210.4356 & $-$20.9830 & 0 & 1.31 & 0.13 & 0.20 & 1.08 \\
HIP 26397       & 174.0618 & +01.5808 & 0 & 0.64 & 0.59 & 0.34 & 1.31 \\
HIP 28881       & 164.9727 & +12.8935 & 0 & 0.24 & 0.18 & 0.37 & 0.83 \\
HIP 29276       & 263.3029 & $-$27.6837 & 0 & 0.70 & 0.17 & 0.17 & 1.24 \\
HIP 31766       & 210.0349 & $-$02.1105 & 3 & 0.90 & 0.61 & 0.45 & 1.66 \\
HIP 32067       & 206.2096 & +00.7982 & 8 & 1.28 & 0.91 & 0.51 & 1.04 \\
HIP 34536       & 224.1685 & $-$00.7784 & 2 & 0.51 & 0.65 & 0.45 & 1.95 \\
HIP 38430       & 243.1553 & +00.3630 & 9 & 1.10 & 0.86 & 0.45 & 1.16 \\
HIP 62322       & 302.4492 & $-$05.2412 & 0 & 0.32 & 0.19 & 0.44 & 1.12 \\
HIP 72510       & 318.7681 & +02.7685 & 0 & 0.94 & 0.43 & 0.33 & 0.93 \\
HIP 75095       & 322.6802 & +00.9060 &0 & 0.40 & 0.26 & 0.52 & 1.03 \\
HIP 77391       & 330.4212 & +04.5928 & 1 & 1.44 & 0.48 & 0.56 & 1.13 \\
HIP 78401$^*$       & 350.0969 & +22.4904 & 0 & 0.57 & 0.15 & 0.34 & 1.06 \\
HIP 81377$^*$       & 006.2812 & +23.5877 & 5 & 0.72 & 0.60 & 0.57 & 1.16 \\
HIP 82171       & 329.9790 & $-$08.4736 & 0  & 1.04 & 0.30 & 0.26 & 1.46 \\
HIP 88652       & 015.1187 & +03.3349 & 0 & 3.00 & 0.28 & 0.29 & 1.35 \\
HIP 92865       & 041.7070 & +03.3784 & 0 & 0.31 & 0.25 & 0.31 & 1.97 \\
HIP 97796$^*$       & 056.4824 & $-$04.3314 & 0 & 1.02 & 0.36 & 0.37 & 1.00 \\
\bd     & 082.4100 & +02.3254 & 0 & 1.00 & 0.33 & 1.05 & 1.19 \\
HIP 114990      & 112.8862 & +03.0998 & 1 & 0.94 & 0.43 & 0.47 & 0.88 \\
\hline                                             
\end{tabular}
\tablefoot{For the 4 bow shocks indicated with stars, the limits were calculated for spatial emission profiles obtained from the WISE infrared emission intensity; for the remaining 23 bow shocks the limits were calculated for a point-like source at the position of the runaway star; see text for details. All bow shock candidates are listed in the \eboss. We calculated upper limits in four logarithmically equally spaced energy bins covering the energy range from 100\,MeV to 300\,GeV. $l$ and $b$ denote the Galactic coordinates of the star. The TS values were calculated assuming a power-law spectrum with photon index of 2 over the whole energy range. \eflux~corresponds to the integral energy flux upper limit within the energy range provided (GeV) assuming a power-law spectrum of gamma-ray emission with photon index $\alpha$=2 within this energy band.}
\end{table*}

\begin{figure}
  \resizebox{\hsize}{!}{\includegraphics{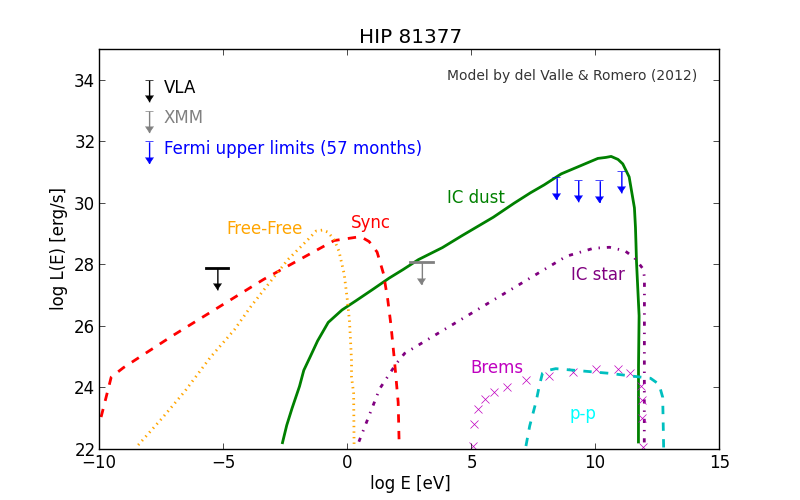}}
  \caption{Upper limits on gamma-ray emission for \zetaoph~in comparison to a model by \cite{Valle_model_zeta}. Also shown are VLA upper limits \citep{vla_skysurvey} and theoretical XMM upper limits \citep{hasinger_xmm_ul_2001}.}
  \label{Fig:spec_HIP81377_src_A}
\end{figure}

Papers reporting model predictions for the high-energy regime have been published for three of the bow shock candidates. 
The most promising candidate for detecting HE emission from a stellar bow shock is \zetaoph~(HIP 81377). \cite{Valle_model_zeta} present a model calculation and conclude that the emission might be detectable at gamma-ray energies. Our upper-limit calculations, based on 57 months of \lat~observations, are lower by a factor of about 5 in the 5.5 - 41\,GeV energy band and therefore severely constrain some of the assumptions in this model as shown in Fig.\,\ref{Fig:spec_HIP81377_src_A}.

\begin{figure}
  \resizebox{\hsize}{!}{\includegraphics{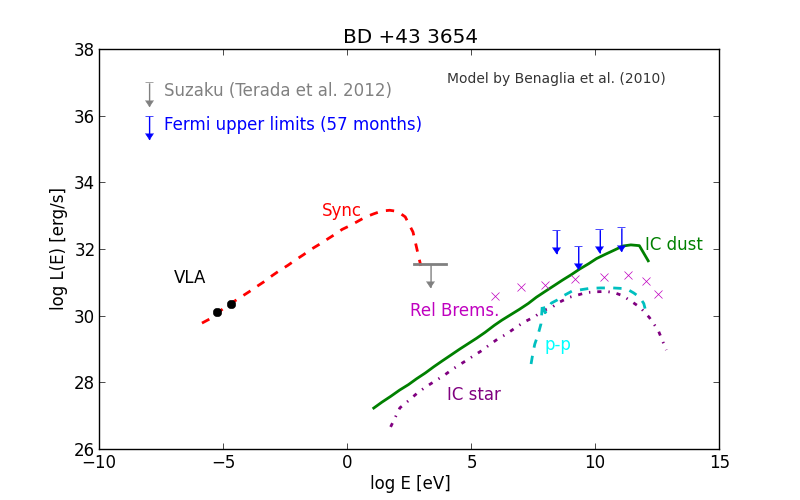}}
  \caption{Upper limits on gamma-ray emission for \bd~in comparison to a model by \cite{bd_benaglia}. In addition, the VLA detection \citep{bd_benaglia} and the \textit{Suzaku} upper limit \citep{terada_bd} are shown.}
  \label{Fig:spec_bd}
\end{figure}

The second candidate is \bd. \cite{bd_benaglia} published the detection of non-thermal radio emission from \bd~and modelled the emission, inferring that the spectral energy distribution extends up to TeV energies. The main contribution in the HE regime is from photons that are created via Compton-upscattering of infrared photons originating in the swept-up dust. \cite{terada_bd} published upper limits on the X-ray emission from \textit{Suzaku}. The model prediction for \bd, the \textit{Suzaku} upper limit, and our upper limits are shown in Fig.\,\ref{Fig:spec_bd}. The gamma-ray limits in the most constraining bin are only about a factor of 4 above the model predictions. 

\begin{figure}
  \resizebox{\hsize}{!}{\includegraphics{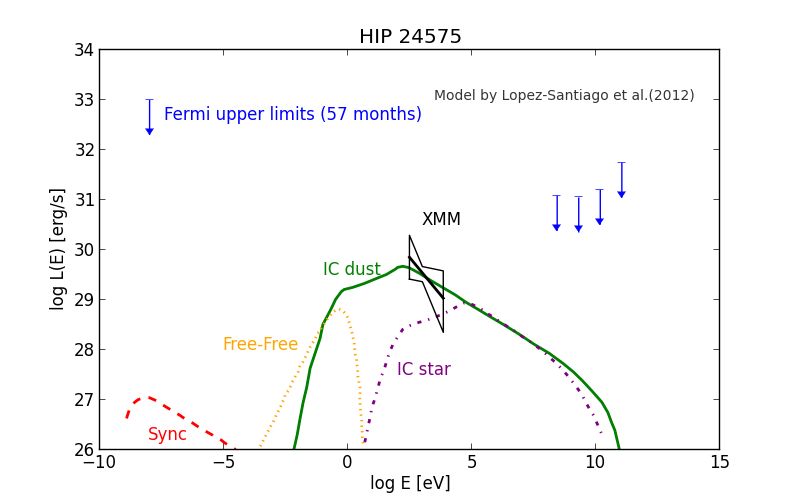}}
  \caption{Upper limits on gamma-ray emission for HIP 24575 in comparison to a model and the XMM detection by \cite{ae_aurigae_LopezSantiago}.}
  \label{Fig:spec_hip24575}
\end{figure}

The third candidate is AE Aurigae (HIP 24575), which is the only stellar bow shock detected in X-rays \citep{ae_aurigae_LopezSantiago}. The model calculation by \cite{ae_aurigae_LopezSantiago} following \cite{Valle_model_zeta} results in a much lower peak energy for the inverse Compton component from the dust photons than for other bow shocks. In this case our upper limits, shown in Fig.\,\ref{Fig:spec_hip24575}, are not constraining. 

\section{Conclusions}
We performed the first systematic study to search for high-energy gamma-ray emission from the bow shocks of runaway stars collected in the \eboss. There is no evidence for high-energy gamma-ray emission in any of the cases. The existing model is challenged by the presented upper limits for one of the bow shock candidates. The spectral energy distributions for the non-thermal emission of bow shocks, computed by \cite{Valle_model_zeta}, mainly depend on the assumptions for the particle acceleration, the magnetic field, and the dust emission. 

One way to interpret our results is that the particle acceleration is not efficient enough, and the maximum energies of the accelerated electrons are less than predicted or that the photon density provided by the dust is lower. Another explanation is that the magnetic fields in these systems are not as turbulent as in other non-thermal emitters like pulsar wind nebulae and supernova remnants \citep[as also pointed out in][]{terada_bd}.  

Further analyses with more data and improved event selection and calibration leading to an improved sensitivity of the LAT \citep[as e.g. outlined in][]{2013arXiv1304.5456B} might constrain the model predictions for \bd. Also, long exposures with current ground-based Cherenkov telescope systems and observations with future Cherenkov telescopes like the Cherenkov Telescope Array \citep[see e.g.][]{Hinton20131}, which will be more sensitive than the \lat~above $\sim$100\,GeV, will be able to test the predictions.

Recently, \cite{2014arXiv1401.3255D} have shown that runaway massive stars can be variable gamma-ray sources on time scales of one to a few years with significant intensity variations between the high and the low states. The variability time scale depends on the size of the density inhomogeneities of the traversed ambient gas and the stellar velocity. A dedicated search for such time variations might help improve the sensitivity of future bow shock searches.

The upper limits presented in this first systematic search provide important constraints on the nature of particle acceleration processes in bow shocks and the environment in which they happen. It therefore helps to improve future emission models for these objects. \\

\textbf{Acknowledgements}

The \textit{Fermi}~LAT Collaboration acknowledges generous ongoing support
from a number of agencies and institutes that have supported both the
development and the operation of the LAT, as well as scientific data analysis.
These include the National Aeronautics and Space Administration and the
Department of Energy in the United States, the Commissariat \`a l'Energie Atomique,
and the Centre National de la Recherche Scientifique / Institut National de Physique
Nucl\'eaire et de Physique des Particules in France, the Agenzia Spaziale Italiana
and the Istituto Nazionale di Fisica Nucleare in Italy, the Ministry of Education,
Culture, Sports, Science and Technology (MEXT), High Energy Accelerator Research
Organization (KEK), and Japan Aerospace Exploration Agency (JAXA) in Japan, and
the K.~A.~Wallenberg Foundation, the Swedish Research Council and the
Swedish National Space Board in Sweden.

Additional support for science analysis during the operations phase is gratefully
acknowledged from the Istituto Nazionale di Astrofisica in Italy and the Centre National d'\'Etudes Spatiales in France.\\

This publication makes use of data products from the Wide-field Infrared Survey Explorer, which is a joint project of the University of California, Los Angeles, and the Jet Propulsion Laboratory/California Institute of Technology, funded by the National Aeronautics and Space Administration. \\

This research has made use of the SIMBAD database, operated at the CDS, Strasbourg, France.

\bibliographystyle{aa}
\bibliography{bowshock_references}

\end{document}